\documentclass{osa-article}
\journal{oe}

\usepackage{bm}

\usepackage{comment}

\begin{document}

\title{First-principles method for nonlinear light propagation at oblique incidence}

\author{Mitsuharu Uemoto,\authormark{1,*} and Kazuhiro Yabana\authormark{2}}

\address{\authormark{1}
Department of Electrical and Electronic Engineering, Graduate School of Engineering, Kobe University, 1-1 Rokkodai-cho, Nada-ku, Kobe 651-8501, Japan
\\
\authormark{2}
Center for Computational Sciences, University of Tsukuba, 1-1-1 Tennodai, Tsukuba, Ibaraki 305-8577, Japan
}

\email{\authormark{*}uemoto@eedept.kobe-u.ac.jp} 


\begin{abstract}
We have developed a computational method to describe the nonlinear light propagation of an intense and ultrashort  
pulse at oblique incidence on a flat surface. 
In the method, coupled equations of macroscopic light propagation and microscopic electron dynamics are simultaneously solved using a multiscale modeling.
The microscopic electronic motion is described by first-principles time-dependent density functional theory.
The macroscopic Maxwell equations that describe oblique light propagation are transformed into one-dimensional wave equations.
As an illustration of the method, light propagation at oblique incidence on a silicon thin film is presented.
\end{abstract}

\section{\label{sec:intro} Introduction}

Light propagation at oblique incidence is one of the fundamental subjects in macroscopic electromagnetism. 
In linear optics, the reflection, refraction, and transmission of $s$- and $p$-polarized light at a flat surface are described by Fresnel's equations and this is used as a basic method to measure optical constants.
An incidence at Brewster's angle is used to avoid the reflection in many optical systems \cite{hecht2002optics}.
A Kretschmann configuration is used for the generation of evanescent fields and surface plasmon polaritons \cite{gwon2010spectral}.
In nonlinear optics, the control of incident angle is essential to achieve phase matching \cite{boyd2003nonlinear}.
Recently, measurements of highly nonlinear and ultrafast electronic dynamics in solids have been attracting interest 
\cite{schultze2013,lucchini2016attosecond,sommer2016attosecond,ciappina2017}.
In such experiments, measurements are often carried out at oblique incidence.
For example, high-harmonic generation from solids at oblique incidence provide useful information on the mechanism and opportunity to control the spectra \cite{vampa2018hhgbackward}.

Although the theoretical description of light propagation at oblique incidence on a flat surface is well documented 
in textbooks of macroscopic electromagnetism \cite{hecht2002optics}, it becomes a research topic when 
the light-matter interaction accompanies nonlinear and/or nonlocal effects.
For example, for oblique-incidence irradiation on periodic nano-structures such as metamaterials or metasurfaces, several theoretical methods have been developed recently including  a field transformation method \cite{veysoglu1993}, a split-field method \cite{roden1998,harms1994implementation,baida2009split} and an iterative method \cite{valuev2008iterative}.
For nonlinear light propagation, constitutive relations with nonlinear optical susceptibilities of second and third orders may be used when the nonlinear effect is sufficiently weak and can be treated perturbatively \cite{varin2018}. 
However, for a strong field that does not allow perturbative description, it is necessary to solve coupled equations 
of macroscopic light propagation and microscopic dynamics of electrons and ions.

Recently, multiscale methods have been developed for nonlinear light propagation that cannot be described using constitutive relations.
In one such method, macroscopic Maxwell equations for light propagation and microscopic quantum-mechanical equations for electronic dynamics are solved simultaneously.
For example, the microscopic electronic dynamics are described by the quantum Liouville equation for the density matrix in the Maxwell-Bloch theory \cite{kira2011sbe}, and those by the time-dependent Kohn-Sham (TDKS) equation for the electronic orbitals in the Maxwell-TDDFT (time-dependent density functional theory) \cite{yabana2012time}.
Since these methods do not use any constitutive relations, \color{blue} they are applicable \color{black} to a propagation with extreme nonlinearity such as high-harmonic generation from thin films \cite{floss2018hhg}, the ultrafast modulation of optical properties by intense pulsed light \cite{lucchini2016attosecond}, and initial stages of laser processing \cite{sato2015propagation}.
However, multiscale calculations have mostly been carried out for the one-dimensional propagation of a plane electromagnetic field at normal incidence where the light propagation can be described by a one-dimensional wave equation. Since microscopic calculations of electron dynamics are computationally expensive, it has been difficult to apply the method to problems with multi-dimensional light propagation.

This paper is intended as a report of a computational method to describe light propagation at oblique incidence in the multiscale Maxwell-TDDFT method.
In this method, the equation to describe the light propagation at oblique incidence is transformed into a one-dimensional wave equation.
Therefore, the computational cost is similar to that in the case of the light propagation at normal incidence. 
Because of the discontinuity of the vector potential at the surface, however, careful implementation is required to obtain an accurate solution.
We note that a method employing the one-dimensional wave-propagation equation has been 
developed for a propagation in linear media at oblique incidence \cite{zhang2010}.
In the field of plasma physics where the microscopic dynamics are described by classical theory such as the Boltzmann equation, a method to describe nonlinear light propagation at oblique incidence has also been developed \cite{liu2018plasma}.

The paper is organized as follows. 
In Sec.~\ref{sec:theory}, we derive a one-dimensional equation that will be used to calculate the light propagation at oblique incidence. 
Then a numerical implementation of the multiscale Maxwell-TDDFT method is explained.
In Sec.~\ref{sec:result}, we show some results of oblique-incidence calculations using the multiscale Maxwell-TDDFT method, taking the irradiations of a flat silicon surface as an example.
Finally, a summary is given in Sec.~\ref{sec:summary}.

\section{\label{sec:theory} Theory} 

\subsection{Multiscale Maxwell-TDDFT method}

We first explain the formalism of the multiscale Maxwell-TDDFT method \cite{yabana2012time} for the case of general three-dimensional light propagation.
In the method, electromagnetic fields that describe light propagation are expressed using a vector potential, ${\bm A}({\bm R},t)$, where we denote the coordinate as ${\bm R}$ for macroscopic quantities.
The Maxwell equation is given by
\begin{equation}
\nabla_{\bm R} \times \nabla_{\bm R} \times {\bm A}({\bm R},t) + \frac{1}{c^2} \frac{\partial^2}{\partial t^2} {\bm A}({\bm R},t)
= \frac{4\pi}{c} {\bm J}({\bm R},t),
\label{eq:Maxwell_3d}
\end{equation}
where ${\bm J}({\bm R},t)$ is the macroscopic electric current density. 

To relate ${\bm J}$ with ${\bm A}$, a local and linear constitutive relation is assumed
in ordinary macroscopic electromagnetism,
\begin{equation}
J_{\alpha}({\bm R},t) = \sum_{\beta} \int^t dt' \sigma_{\alpha\beta}(t-t') \left( -\frac{1}{c} \frac{\partial}{\partial t} A_{\beta}({\bm R},t') \right),
\label{eq:conductivity}
\end{equation}
where $\alpha, \beta$ represent Cartesian indices, $x, y, \text{or} z$, and $\sigma_{\alpha\beta}(t-t')$ is the linear conductivity tensor.
In the multiscale Maxwell-TDDFT method, we use TDDFT \cite{runge1984density} in the time-domain calculation \cite{yabana1996time,bertsch2000} to relate ${\bm J}$ with ${\bm A}$. 
We retain the locality in the relation, assuming that the electric current density at ${\bm R}$, ${\bm J}({\bm R},t)$, is determined by the vector potential at the same position, ${\bm A}({\bm R},t)$.
We also assume that the macroscopic field ${\bm A}({\bm R},t)$ is sufficiently smooth at the atomic scale so that the microscopic electronic motion can be treated in the dipole approximation.
Namely, at each position ${\bm R}$, we consider a microscopic electronic system that is infinitely periodic and is subject to the spatially uniform electric field given by ${\bm A}({\bm R},t)$.

In practical calculations, we introduce a Bloch orbital $u_{{\bm R}, b{\bm k}}({\bm r},t)$ to describe the microscopic electronic motion at ${\bm R}$.
Here, $b$ is the band index and ${\bm k}$ is the Bloch wavenumber.
A coordinate ${\bm r}$ is introduced to describe microscopic electronic motion.
The Bloch orbital satisfies the following TDKS equation;
\begin{align}
i\hbar  \frac{\partial}{\partial t}
u_{{\bm R}, b{\bm k}}({\bm r}, t)
=&
\left[
\frac{1}{2m} \left(
-i\hbar \nabla_{\bm r}
+ \hbar{\bm k}
+ \frac{e}{c}{\bm A}({\bm R}, t)
\right)^2
\right.
\notag \\ &
\color{blue}
+ \hat{V}_{\rm ion}
+ \hat{V}_{\mathrm{H}, {\bm R}}({\bm r}, t)
+ \hat{V}_{\mathrm{xc}, {\bm R}}({\bm r}, t)
\color{black}
\bigg]
u_{{\bm R}, b {\bm k}}({\bm r}, t)
\label{eq:tdks}
\;,
\end{align}
where $\hat{V}_{\rm ion}$, $\hat{V}_{\mathrm{H}, {\bm R}}$ and $\hat{V}_{\mathrm{xc}, {\bm R}}$ are the electron-ion, Hartree, and exchange-correlation potentials, respectively.
We note that the coordinate ${\bm R}$ in the above equation is treated as a parameter
independent of the coordinate ${\bm r}$.

From the Bloch orbitals, the electric current density at ${\bm R}$ is obtained as 
\begin{align}
{\bm J}({\bm R}, t)
=&
\frac{-e}{\Omega_\mathrm{cell}}
\sum_{\bm k}^\mathrm{BZ}
\sum_{b}^\mathrm{occ}
\color{blue}
w_{\bm k}
\color{black}
\int_{\Omega_\mathrm{cell}}
\mathrm{d}{\bm r} \; 
u^*_{{\bm R}, b {\bm k}}({\bm r}, t)
\notag \\ &
\left(
-i\hbar \nabla_{\bm r}
+ \hbar {\bm k}
+ \frac{e}{c}{\bm A}({\bm R}, t)
\color{blue}
- \frac{i}{\hbar} \left[ \hat{V}_{\rm ion}, {\bm r}\right]
\color{black}
\right)
u_{{\bm R}, b {\bm k}}({\bm r}, t)
\label{eq:current}
\;,
\end{align}
where $\Omega_\mathrm{cell}$ is the volume of the unit cell, and the summations of ${\bm k}$ and $b$ are performed over the entire Brillouin zone (BZ) and occupied bands (occ), respectively;
\color{blue}
$w_{\bm k}$ is the weight of $k$ point.
The term $[ \hat{V}_{\rm ion}, {\bm r} ]$ appears when we use a nonlocal pseudopotential.
\color{black}

Equations~(\ref{eq:tdks}) and (\ref{eq:current}) without the coordinate index ${\bm R}$ are known as the basic equations of real-time TDDFT in crystalline solids and have been utilized for many purposes such as dielectric functions \cite{bertsch2000}, nonlinear susceptibilities \cite{uemoto2019nonlinear}, high-harmonic generation \cite{otobe2012hhg}, and energy transfer from a light pulse to electrons \cite{yamada2019energy}.

Solving Eqs.~(\ref{eq:tdks}), (\ref{eq:current}), and (\ref{eq:Maxwell_3d}) simultaneously,
we can describe macroscopic light propagation as well as microscopic electron dynamics.
In practical calculations, we introduce a grid system in the macroscopic coordinate ${\bm R}$ 
to solve Eq.~(\ref{eq:Maxwell_3d}).
The Bloch orbital $u_{{\bm R}, b{\bm k}}({\bm r},t)$ is prepared and the TDKS equation~(\ref{eq:tdks}) is solved at each grid point ${\bm R}$ inside the medium.
To start the time-evolution calculation, the Bloch orbital at every grid point is set to the ground state in the static density functional theory, while an incident pulse is set in the vector potential in the vacuum area.
We note that, when the electric field is sufficiently weak, the electric current given by Eq.~(\ref{eq:current}) is related to the vector potential that is used in Eq.~(\ref{eq:tdks}) by the linear constitutive relation of Eq.~(\ref{eq:conductivity}).
Therefore, the multiscale Maxwell-TDDFT method results into an ordinary macroscopic Maxwell equation with the linear constitutive relation given by the linear response TDDFT \cite{bertsch2000}.

Since it is necessary to solve the TDKS equation (\ref{eq:tdks}) at all grid points of ${\bm R}$ in the medium simultaneously, it becomes a large-scale calculation even for one-dimensional 
light propagation.
Because of the high computational cost, applications of the multiscale Maxwell-TDDFT method 
have so far been limited to cases of pulsed light irradiated normally on flat surfaces or thin films for which the Maxwell equation (\ref{eq:tdks}) is a one-dimensional wave equation 
\cite{sato2015propagation,lucchini2016attosecond,sommer2016attosecond,uemoto2021graphite,yamada2019cp}.

\subsection{Maxwell equations at oblique incidence}

\begin{figure}[h!]
\centering
\includegraphics[width=0.45\textwidth]{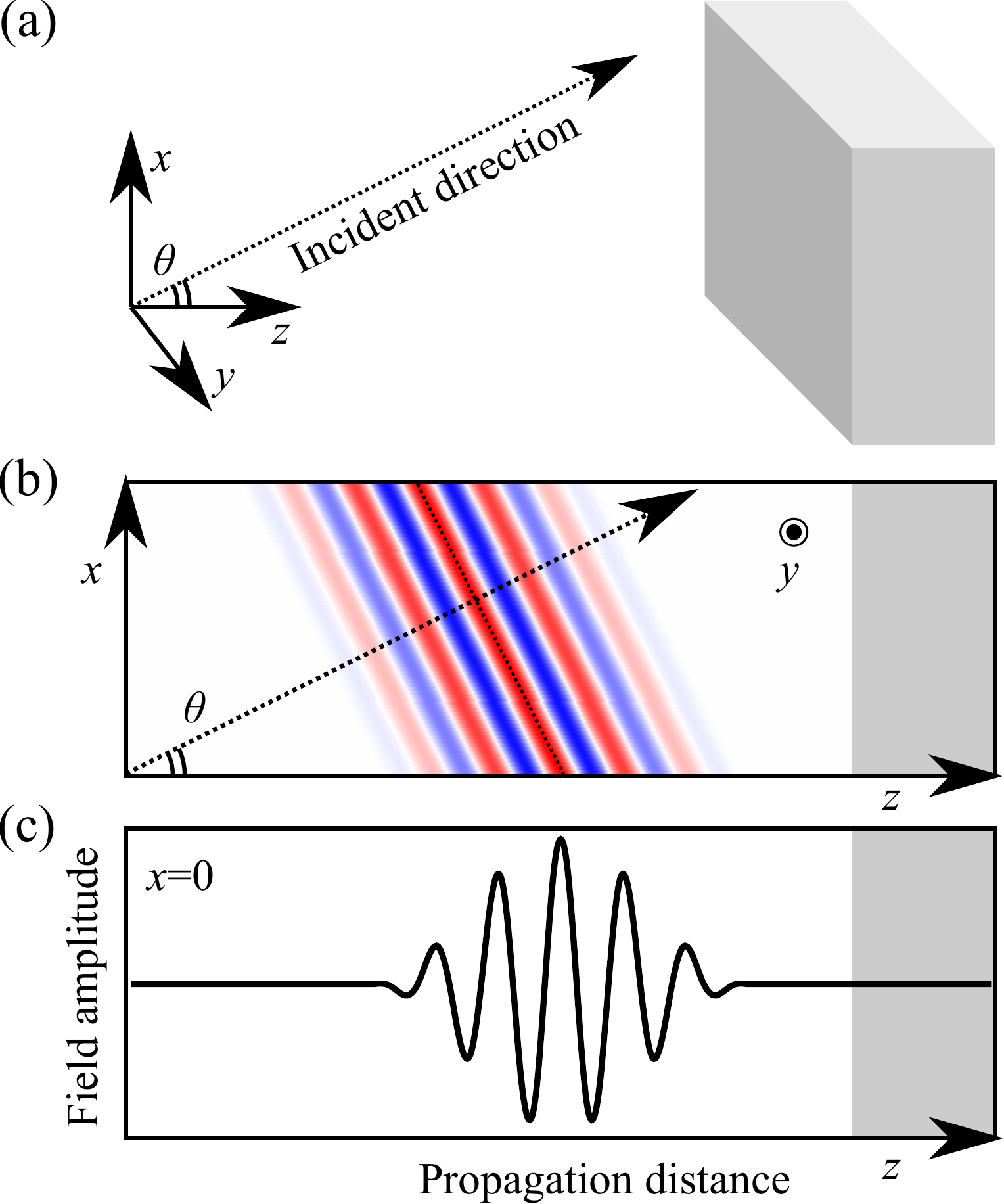}
\caption{ 
\label{fig:illust}
\color{blue}
(a) Setup of the system and coordinates for the incident angle  $\theta$.
(b) The incident electric field is shown in the $XZ$-plane.
(c) Electric field for the cross section along the $Z$-axis. 
}
\end{figure}

We next show that the Maxwell equations for the light propagation at oblique incidence can be expressed as a one-dimensional wave equation.
We consider the irradiation of a flat surface by a plane wave of pulsed light with the incident angle of $\theta$, as shown in Fig.~\ref{fig:illust}(a).
We express the macroscopic coordinate as ${\bm R} = (X,Y,Z)$.
Uniform matter is in the $Z>0$ region with the surface at the $Z=0$ plane.
For the incident plane wave, we choose the $XZ$-plane as the incident plane.
The electromagnetic fields are uniform in the $Y$-direction so that the vector potential is independent of the $Y$-coordinate.

From the translational symmetry, we find that the vector potential and electric current density have the following dependence on the coordinates $X$, $Z$, and time $t$:
\begin{align}
{\bm A}({\bm R},t) &= {\bm a} \left( Z, t - \frac{X \sin\theta}{c} \right),
\label{eq:transform_a}
\end{align}
\begin{align}
{\bm J}({\bm R},t) &= {\bm j} \left( Z, t-\frac{X\sin\theta}{c} \right).
\label{eq:transform_j}
\end{align}
where ${\bm a}$ and ${\bm j}$ are functions depending on the two variables $Z$ and $t$.
We note that this relation holds for pulses irrespective of their intensity and polarization.

Using ${\bm a}$ and ${\bm j}$, the Maxwell equation (\ref{eq:Maxwell_3d}) is written for each Cartesian components as follows. For $a_Y$, we have
\begin{equation}
\frac{\cos^2\theta}{c^2}  \frac{\partial^2}{\partial t^2} a_Y
- \frac{\partial^2}{\partial Z^2} a_Y = \frac{4\pi}{c} j_Y.
\label{eq:Maxwell_y}
\end{equation}
For $a_X$ and $a_Z$, we have the following coupled equations:
\begin{equation}
\frac{1}{c^2} \frac{\partial^2}{\partial t^2} a_X - \frac{\sin\theta}{c} \frac{\partial^2}{\partial t \partial Z} a_Z
- \frac{\partial^2}{\partial Z^2} a_X = \frac{4\pi}{c} j_X,
\label{eq:Maxwell_xz1}
\end{equation}
\begin{equation}
\frac{\cos^2\theta}{c^2} \frac{\partial^2}{\partial t^2} a_Z
- \frac{\sin\theta}{c} \frac{\partial^2}{\partial t \partial Z} a_X = \frac{4\pi}{c} j_Z.
\label{eq:Maxwell_xz2}
\end{equation}
In general, Eqs.~(\ref{eq:Maxwell_y}), (\ref{eq:Maxwell_xz1}), and (\ref{eq:Maxwell_xz2})
are coupled and must be solved simultaneously.
For an isotropic linear medium, Eq. (\ref{eq:Maxwell_y}) describes the propagation of $s$-polarized light, while Eqs. (\ref{eq:Maxwell_xz1}) and (\ref{eq:Maxwell_xz2}) describe the propagation of $p$-polarized light.

For the coupled equations of (\ref{eq:Maxwell_xz1}) and (\ref{eq:Maxwell_xz2}), we further rewrite them as following.
First, we integrate Eq. (\ref{eq:Maxwell_xz2}) over time to obtain
\begin{equation}
\frac{\cos^2\theta}{c^2} \frac{\partial}{\partial t} a_Z
- \frac{\sin\theta}{c} \frac{\partial}{\partial Z} a_X = \frac{4\pi}{c} p_Z,
\label{eq:Maxwell_xz21}
\end{equation}
where we introduced the polarization $p_Z(Z,t)$ by
\begin{equation}
p_Z(Z,t) = \int^t dt' j_Z(Z,t').
\end{equation}
Differentiating Eq. (\ref{eq:Maxwell_xz21}) with respect to $Z$ and combining it with Eq. (\ref{eq:Maxwell_xz1}), 
we have
\begin{equation}
\frac{\cos^2\theta}{c^2} \frac{\partial^2}{\partial t^2}a_X - \frac{\partial^2}{\partial Z^2} a_X
= \frac{4\pi \cos^2\theta}{c} j_X + 4\pi \sin\theta \frac{\partial}{\partial Z} p_Z.
\label{eq:Maxwell_p}
\end{equation}
We will use this equation to calculate the time evolution of $a_X$.
Once $a_X$ is obtained, $a_Z$ may be obtained by integrating Eq.~(\ref{eq:Maxwell_xz21}) over time.
Equations.~(\ref{eq:Maxwell_y}), (\ref{eq:Maxwell_xz21}), and (\ref{eq:Maxwell_p}), which are partial differential equations of $Z$ and $t$, are the basic equations
to be solved numerically to calculate the light propagation at oblique incidence.

\subsection{\label{sec:bc} Boundary conditions}

To carry out numerical calculations, it is important to clarify the behavior of the vector potential at the interface, $Z=0$.
In this subsection, we will examine the boundary conditions for each component of the vector potential.
To confirm that the boundary conditions are consistent with the ordinary results, 
we examine the reflection and transmission through the surface for an isotropic and linear medium.

We first examine the boundary condition for $a_Y$.
In Eq. (\ref{eq:Maxwell_y}), there appear $a_Y$ and $j_Y$.
The current $j_Y$ is discontinuous at $Z=0$. Accordingly, $(\partial^2/\partial Z^2)a_Y$ must also be discontinuous at $Z=0$. This fact indicates that both $a_Y$ and $(\partial/\partial Z) a_Y$ are continuous at $Z=0$. We consider the consequence of these boundary conditions for an incident pulse of $s$-polarization on an isotropic and linear medium characterized by the dielectric constant $\epsilon$.
The following form is assumed for the solution:
\begin{equation}
a_Y(Z,t) = \left\{
\begin{array}{ll}
I_s e^{ikZ-i\omega t} + R_s e^{-ikZ-i\omega t} & (Z<0) \\
T_s e^{ik'Z-i\omega t} & (Z>0),
\end{array} \right.
\end{equation}
where the wave numbers $k$ and $k'$ are given by $k=(\omega/c)\cos\theta$ and $k'=(\omega/c) \sqrt{\epsilon-\sin^2\theta}=(\omega/c)n \cos\theta'$, respectively.
The refraction index $n$ is introduced by $n=\sqrt{\epsilon}$.
The angle $\theta'$ is the refraction angle and satisfies Snell's law,
\begin{equation}
\frac{\sin \theta'}{\sin\theta} = \frac{1}{n}.
\end{equation}

The coefficients $R_s$ and $T_s$ are determined by the boundary condition that both $a_Y$ and $(\partial/\partial Z)a_Y$ be continuous at $Z=0$.
Then, we have
\begin{equation}
I_s + R_s = T_s,
\end{equation}
\begin{equation}
kI_s - kR_s = k'T_s.
\end{equation}
From these relations, we obtain the well-known result for  $s$-polarization,
\begin{equation}
\frac{R_s}{I_s} = \frac{\cos\theta - n \cos\theta'}{\cos\theta + n \cos\theta'},
\end{equation}
\begin{equation}
\frac{T_s}{I_s} = \frac{2 \cos\theta}{\cos\theta + n \cos\theta'}.
\end{equation}

We next consider boundary conditions for $a_X$ and $a_Z$ components.
In Eq.~(\ref{eq:Maxwell_p}), $(\partial/\partial Z) p_Z$ on the right-hand side is divergent at $Z=0$ since $p_Z$ is discontinuous.
We integrate both sides of Eq.~(\ref{eq:Maxwell_p}) over a small region of $Z$ including $Z=0$, $[-\delta, +\delta]$ with an infinitesimally small $\delta$.
Then, we have
\begin{equation}
\left. - \frac{\partial}{\partial Z} a_X \right \vert^{+\delta}_{-\delta}
= 4\pi \sin\theta p_Z(Z=+\delta, t).
\label{eq:bc_ax}
\end{equation}
From this relation, we find that $(\partial/\partial Z) a_X$ is discontinuous and,
then, $a_X$ is continuous at $Z=0$.

To find the boundary condition for $a_Z$, we evaluate Eq. (\ref{eq:bc_ax}) at $Z=\pm \delta$ and combine it with Eq. (\ref{eq:Maxwell_xz21}) to obtain
\begin{equation}
\left. -\frac{1}{c} \frac{\partial}{\partial t} a_Z \right\vert^{+\delta}_{-\delta}
+ 4\pi p_Z(Z=+\delta, t) = 0.
\label{eq:bc_axz}
\end{equation}
From this relation, we find that $a_Z$ is discontinuous at $Z=0$.
We note that Eq. (\ref{eq:bc_axz}) is the ordinary continuity condition for the perpendicular component of the electric displacement, $D_Z = E_Z + 4\pi P_Z$.

We consider the case of an incident pulse of $p$-polarization on an isotropic and linear medium.
The solution can be expressed as
\begin{align*}
a_X
=&
\left\{
\begin{aligned}
&
\left(
I_p 
\cos\theta
e^{ikZ}
+
R_p 
\cos\theta 
e^{-ikZ}
\right) e^{- i\omega t}
& (Z<0) \\
&T_p 
\cos\theta' 
e^{ik'Z-i\omega t}
& (Z>0)
\end{aligned} \right.
\\
a_Z
=&
\left\{
\begin{aligned}
&
\left(
-I_p 
\sin\theta
e^{ikZ}
+
R_p 
\sin\theta 
e^{-ikZ}
\right) e^{- i\omega t}
& (Z<0) \\
&-T_p 
\sin\theta' 
e^{ik'Z-i\omega t}
& (Z>0)
\end{aligned} \right.
\;.
\end{align*}
To determine the coefficients $R_p$ and $T_p$, we utilize the continuity of $a_X$ and Eq.~(\ref{eq:bc_axz}).
They provide
\begin{equation}
I_p \cos\theta + R_p \cos\theta = T_p \cos\theta',
\end{equation}
\begin{equation}
I_p \sin\theta - R_p \sin\theta = n^2 T_p \sin\theta'.
\end{equation}
From these relations, we obtain the well-known result for $p$-polarization,
\begin{equation}
\frac{R_p}{I_p} = \frac{\cos\theta' - n \cos\theta}{\cos\theta' + n \cos\theta},
\end{equation}
\begin{equation}
\frac{T_p}{I_p} = \frac{2 \cos\theta}{\cos\theta' + n \cos\theta}.
\end{equation}

\subsection{\label{sec:numerical} Numerical implementation}

\begin{figure}[h!]
\centering
\includegraphics[width=0.65\textwidth]{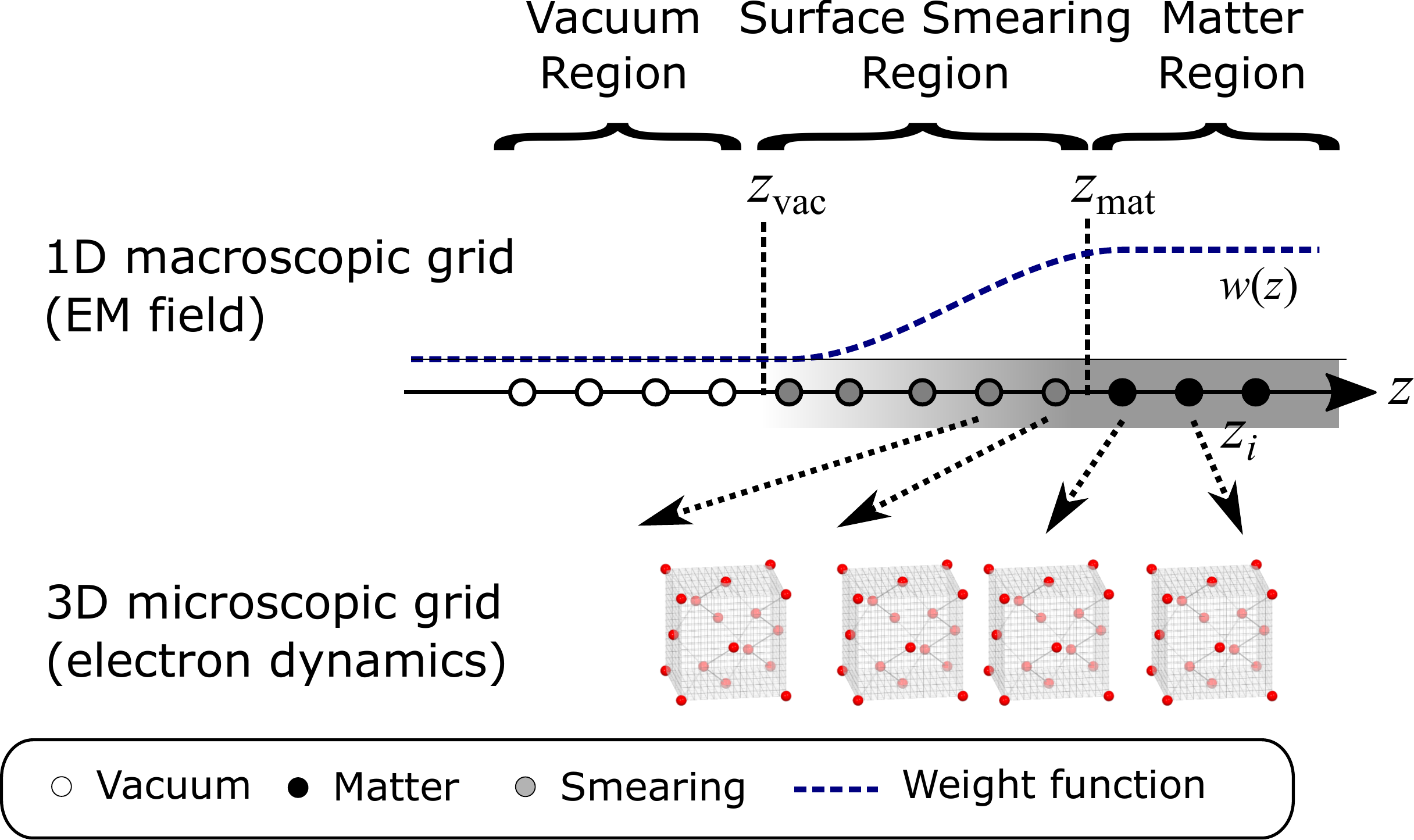}
\caption{ 
\label{fig:multiscale}
Spatial grid systems used in the multiscale Maxwell-TDDFT method.
The smearing region is indicated by gradation.
}
\end{figure}

To achieve a multiscale Maxwell-TDDFT calculation at oblique incidence, we solve the one-dimensional propagation equations
(\ref{eq:Maxwell_y}), (\ref{eq:Maxwell_xz21}), and (\ref{eq:Maxwell_p}) together with the microscopic equations
 (\ref{eq:tdks}) and (\ref{eq:current}).
To solve the one-dimensional propagation equations, we introduce a uniform spatial grid $\{ Z_i \}$ in the $Z$ coordinate.
We solve the propagation equations using finite difference approximation and do not use the boundary conditions at $Z=0$.
As discussed in the previous subsection, there appears a function $(\partial / \partial Z) p_Z$ that diverges at $Z=0$ on the right-hand side of Eq.~(\ref{eq:Maxwell_p}), and such divergence is cancelled by a divergence of the 
$(\partial^2 /\partial Z^2) a_X$ term.
We numerically found that the divergent term induces unphysical oscillations if we naively implement the finite difference approximation.

To avoid this numerical difficulty, we have found it useful to introduce smearing in the vicinity of the
boundary, $[Z_{\rm vac}, Z_{\rm mat}]$, as shown in Fig.~\ref{fig:multiscale}.
In the smearing region, we multiply a weight function $w$ to the electric current density,
\begin{align}
\tilde{\bm J}(Z_i, t)
  =&
  \begin{cases}
  0 & \text{(Vacuum region)} \\
  w(Z_i){\bm J}(Z_i, t) & \text{(Smearing region)}\\
  {\bm J}(Z_i, t) & \text{(Matter region)} 
  \end{cases}
  \;.
\end{align}
The weight function $w$ is a smooth function, changing from 0 in the vacuum region to 1 in the matter region.
In our implementation, we will use the following function that is defined in $[Z_{\rm vac}, Z_{\rm mat}]$:
\begin{align}
    w(Z)
    =&
-2 \left(
\frac{Z - Z_\mathrm{vac}}{Z_\mathrm{mat} - Z_\mathrm{vac}}
\right)^3
+3 \left(
\frac{Z - Z_\mathrm{vac}}{Z_\mathrm{mat} - Z_\mathrm{vac}}
\right)^2
\;,
\end{align}
which satisfies $w(Z_\mathrm{vac})=0$, $w(Z_\mathrm{mat})=1$, and
$w'(Z_\mathrm{vac})=w'(Z_\mathrm{mat})=0$.

To update the vector potential, we adopt the following scheme:
\begin{align}
a_{Y, i}^{(m+1)}
=&
2\left(1- \frac{c^2 \Delta{t}^2}{\Delta{Z}^2 \cos^2 \theta}\right) a_{Y, i}^{(m)} - a_{Y, i}^{(m-1)}
\notag \\ &
+\frac{c^2 \Delta{t}^2}{\Delta{Z}^2 \cos^2 \theta} a_{Y, i+1}^{(m)}
+\frac{c^2 \Delta{t}^2}{\Delta{Z}^2 \cos^2 \theta} a_{Y, i-1}^{(m)}
\notag \\ &
+\frac{4 \pi \Delta{t}^2 c}{\cos^2 \theta} \tilde{J}_{Y, i}^{(m)}
\label{eq:fdtd_oblique1}
\\
a_{X, i}^{(m+1)}
=&
2\left(1- \frac{c^2 \Delta{t}^2}{\Delta{Z}^2 \cos^2 \theta}\right) a_{X, i}^{(m)}
- a_{X, i}^{(m-1)}
\notag \\ &
+ \frac{c^2 \Delta t^2}{\Delta Z^2 \cos^2\theta} a_{X, i+1}^{(m)}
+ \frac{c^2 \Delta t^2}{\Delta Z^2 \cos^2\theta} a_{X, i-1}^{(m)} 
\notag \\ &
+ \frac{2 \pi c ^2 \Delta{t}^2 \sin\theta}{\Delta{Z}\cos^2\theta} \tilde{P}^{(m)}_{Z, i+1}
- \frac{2 \pi c ^2 \Delta{t}^2 \sin\theta}{\Delta{Z}\cos^2\theta} \tilde{P}^{(m)}_{Z, i-1}
\notag \\ &
+4 \pi c \Delta{t}^2 \tilde{J}^{(m)}_{X, i}
\label{eq:fdtd_oblique2}
\\
a_{Z, i}^{(m+1)}
=&
a_{Z, i}^{(m-1)}
+ \frac{c \Delta{t}}{\Delta{Z} \cos^2\theta} a_{Z, i+1}^{(m)}
- \frac{c \Delta{t}}{\Delta{Z} \cos^2\theta} a_{Z, i-1}^{(m)}
\notag \\ &
+ \frac{8 \pi c \Delta{t}}{\cos^2\theta} 
\color{blue} \tilde{P}^{(m)}_{Z, i} 
\;,
\label{eq:fdtd_oblique3}
\end{align}
where the subscript $i$ indicates the spatial index and the superscript $(m)$ indicates the time index, respectively.
The polarization is updated by
\begin{align}
\tilde{{\bm P}}^{(m+1)}_{i} =&\tilde{{\bm P}}^{(m-1)}_{i} + 2 \Delta{t} \tilde{{\bm J}}^{(m)}_{i} .
\end{align}

\begin{figure}[h!]
\centering
\includegraphics[width=0.75\textwidth]{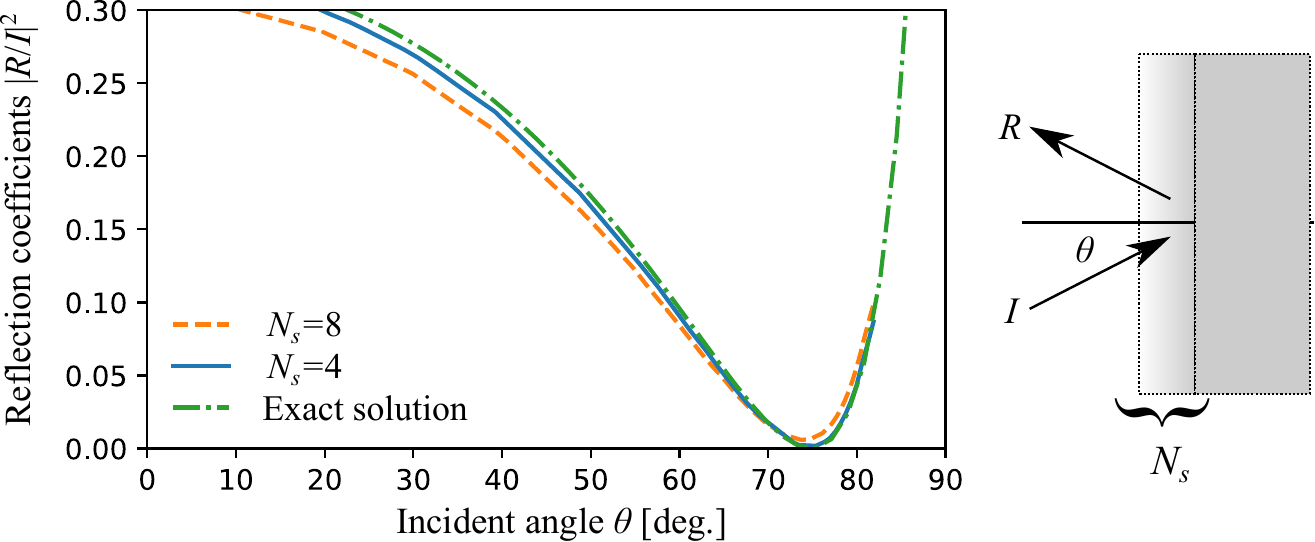}
\caption{ 
\label{fig:reflection}
Reflectance of $p$-polarized light from a surface of a linear medium as a function of the incident angle $\theta$. 
Numerical results with different smearing points, $N_s=4$ and 8, are compared with the exact value.}
\end{figure}

\color{blue}
In Eqs.~(\ref{eq:fdtd_oblique1})-(\ref{eq:fdtd_oblique3}), $1/(\cos\theta)^2$ becomes divergent as $\theta$ approaches $90^\circ$ (grazing incidence), and the method does not work in the limit. This is already evident in the 1D transformed equations  (\ref{eq:Maxwell_y})-(\ref{eq:Maxwell_xz2}) , which are singular as $\theta$ approaches $90^\circ$.
\color{black}

To verify that the smearing method is effective, we performed a test calculation of pulse propagation at oblique incidence on a bulk surface of a linear medium.
We use an incident pulse with the central frequency $\omega_c = 1.55$ eV.
A dielectric function of a classical Drude-Lorentz model,
\begin{equation}
\epsilon(\omega) = 1+\frac{4\pi\alpha}{\omega_i^2 - \omega^2 - i\gamma \omega},
\end{equation}
is used for the response of the medium.
The parameters are set as $\alpha=4$~au and $\omega_0=2$~au, which reasonably describes the dielectric constant of silicon ($\epsilon \approx 13.6$) \cite{ghosh1998handbook}.

In Fig.~\ref{fig:reflection}, we show the reflectance of $p$-polarized light as a function of the incident angle $\theta$.
In the calculation, a grid spacing of $\Delta{Z}=0.53$~nm is used.
The results of calculations using different numbers of grid points $N_s$ for smearing, $N_s=8$ (orange dotted curve) and $N_s=4$ (blue solid curve), are compared with the results of exact Fresnel formula of reflection (green broken curve) \cite{hecht2002optics}:
\begin{align}
|r(\theta)|^2 &=
\left|
\frac{\tan(\theta-\theta')}{\tan(\theta+\theta')} 
\right|^2
\;.
\end{align}
The numerical results are in good agreement with the exact value. The minimal reflectance at the Brewster angle of $\theta \approx 75^\circ$ is accurately described by the numerical calculation with the smearing.
A comparing calculations with $N_s=4$ and $8$, the accuracy is improved by decreasing the smearing points $N_s$.
This indicates that, as expected, the smearing region $N_s \Delta Z$ should be small to obtain on accurate result.
However, we find that a finer grid spacing $\Delta Z$ is required to obtain a stable numerical solution when a small $N_s$ value is adopted.

\color{blue}
Let us discuss the limiting behavior of the smearing method when we use a small smearing area and a small grid spacing.
For simplicity, we restrict ourselves to the linear regime in which our Maxwell-TDDFT formalism reduces to ordinary macroscopic electromagnetism with a dielectric function in TDDFT. In our smearing method, it is possible to improve the description as much as we want by simultaneously decreasing the length of the smearing area and also the grid spacing. In the limit, it is possible to reproduce exact results of boundary-value problem at an interface in macroscopic electromagnetism. Therefore, using our smearing method with sufficiently small smearing area and grid spacing, it is possible to describe phenomena that can be described in ordinary macroscopic electromagnetism such as the Fresnel reflection, as confirmed above, and the surface plasmon at the metal-dielectric interface.
\color{black}

\section{Illustrative calculations}
\label{sec:result}

\begin{figure}[h!]
\centering
\includegraphics[width=0.65\textwidth]{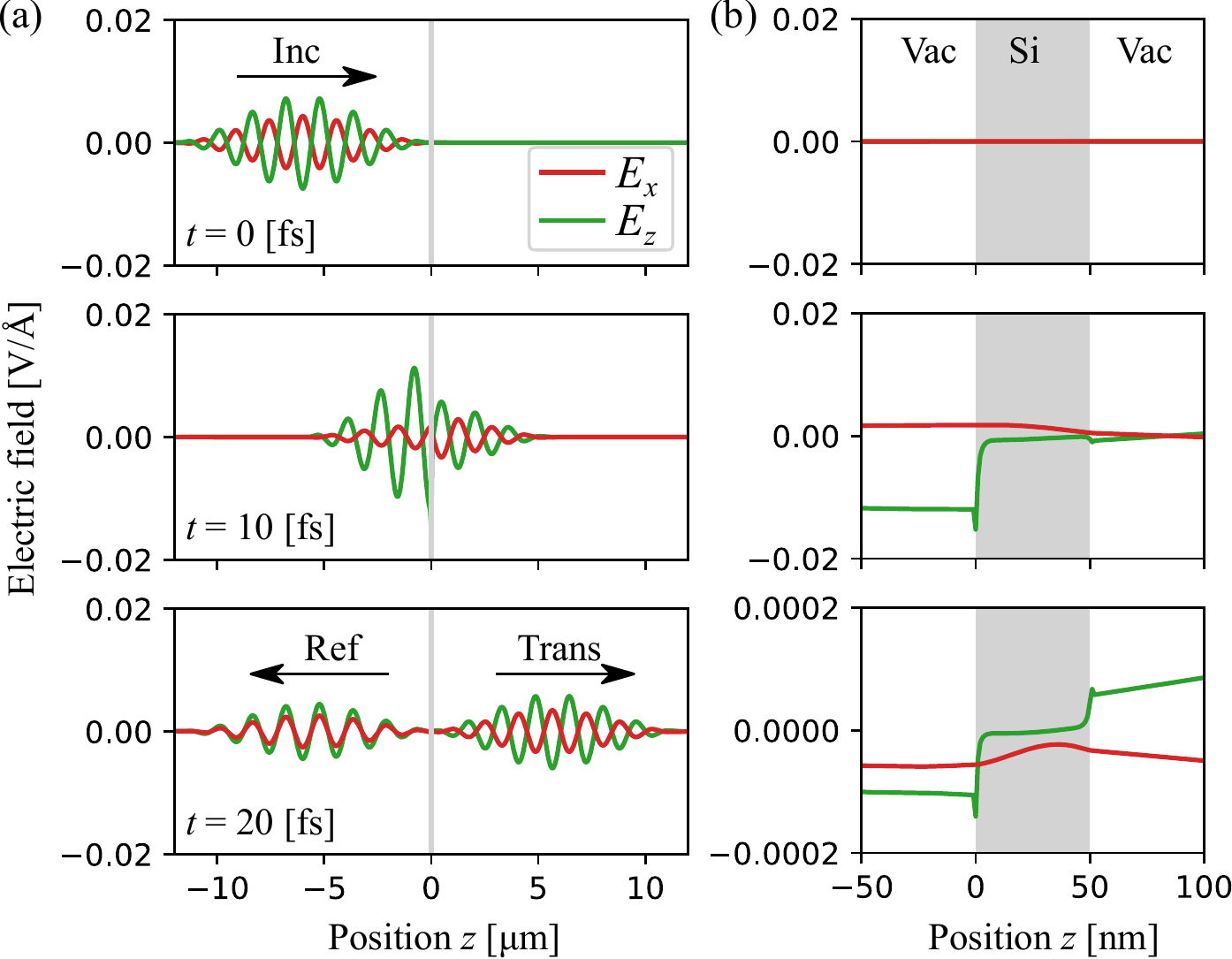}
\caption{ 
\label{fig:field}
(a) The electric field in the multiscale Maxwell-TDDFT calculation is shown at $t=0$ fs (top),
$t=10$ fs (middle), and $t=20$ fs (bottom). $E_X$ is shown by the red line and $E_Z$ by the green line.
The silicon thin film is located within $ 0 < Z < 0.05$ $\mu$m.
(b) The expanded view of the electric field in the thin film region is also shown.
}
\end{figure}

\begin{figure}[h!]
\centering
\includegraphics[width=0.95\textwidth]{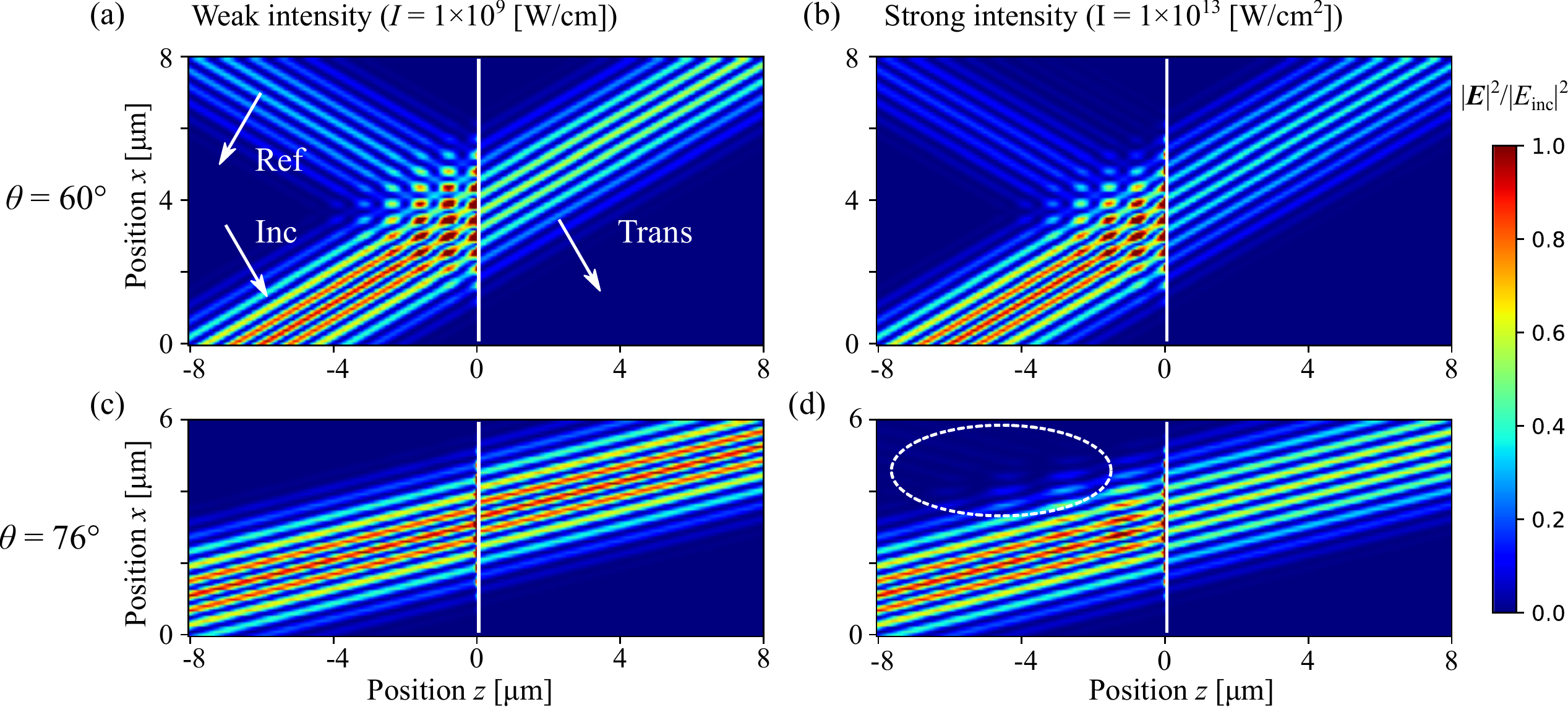}
\caption{\label{fig:profile}
Electric field intensity, $|E(X,Z,t)|^2$ at a fixed time $t$ calculated by the multiscale Maxwell-TDDFT method.
Two incident angles of $\theta = 60^\circ$ and $76^\circ$ (Brewster's angle) , and two incident intensities of 
$I=10^9$ W/cm$^2$ (weak) and $10^{13}$ W/cm$^2$ (strong) are shown in panels (a)-(d).
The magnitude of the field intensity is normalized by the maximum value of the incident pulse.
}
\end{figure}

In this section, we show several calculations of light propagations at oblique incidence on a silicon thin film 
of 50 nm thickness using the multiscale Maxwell-TDDFT method.
For the incident pulse, we use a pulse with a cos-square-shaped envelope with the central frequency
of $\hbar \omega = 1.55$ eV and the full duration of $T_p = 20$ fs. 
We set the $Z=0$ plane as the $(001)$ surface of silicon.
A linear $p$-polarization is assumed for the incident pulse with the incident plane parallel to $(010)$.
Because of the symmetry, there is no electric current density in the $Y$-direction.
Therefore, only two propagation equations  (\ref{eq:Maxwell_xz21}) and (\ref{eq:Maxwell_p}) 
are utilized to describe the propagation.

We have implemented the light propagation at oblique incidence in the SALMON code \cite{noda2019salmon} that has been developed by the authors' group.
In the code, a real-space 3D grid is used to express electron orbitals.
A norm-conserving pseudopotential is used for the electron-ion interaction \cite{troullier1993,kleinman1982}.
Numerical parameters are chosen as follows.
For the propagation of the light pulse, we use a spatial grid size of $\Delta Z = 1.0$~nm so that the number of grid points in the Si film is $N=50$.
We choose the number of smearing points $N_s$=8 for the front and back surfaces.
For the electron dynamics calculation, we use a cubic unit cell containing eight silicon atoms.
Each side of the cell is divided into 16 grid points.
The $k$-space is sampled by $8 \times 8 \times 8$ Monkhorst-Pack grids.
A common time step is used for the light propagation and electron dynamics calculations, $\Delta t = 0.43$~as.

We use the adiabatic local density approximation (ALDA) as the exchange-correlation potential \cite{perdew1981}. 
It is well known that the LDA systematically underestimates the optical gap energy of dielectrics, which results in the overestimation of the index of refraction $n$.  
In the case of bulk Si, ALDA gives $n=4.0$ at the frequency of $1.55$~eV, while the measured value is $n \simeq 3.7$ \cite{ghosh1998handbook}.

Figure \ref{fig:field}(a) shows snapshots of the two components of the electric field, $E_X$ (red) 
and $E_Z$ (green).
The thin film is placed in the $Z=0$ region and appears as a line without thickness at this scale.
The incident angle is chosen at $60^\circ$, which is smaller than Brewster's angle.
The maximum intensity of the pulse is set at $I=10^9$~W/cm$^2$.
At $t = 0$ fs (top), the incident pulse is placed at the left of the thin film and propagates in the positive-$Z$ direction.
At $t=10$ fs (middle), the pulse reaches the film and interacts with it.
At $t=20$ fs (bottom), the pulse is separated into reflected and transmitted waves.
Figure~\ref{fig:field}(b) shows expanded views of the electric field in the vicinity of the thin film.
It can be seen that  $E_X$ is continuous at the surfaces of the film, while $E_Z$ is discontinuous. 
This behavior is consistent with the boundary condition discussed in Sec.~\ref{sec:bc}.

Figure~\ref{fig:profile} shows two-dimensional views of the electric field intensity, $\vert {\bm E}(X,Z,t) \vert^2$, at a certain time $t$ where the electric field is generated by 
${\bm E}(X,Z,t) = -(1/c) (\partial/\partial t) {\bm a}(Z, t-X\sin\theta/c)$.
For the cases of Fig.~\ref{fig:profile}(a) is the case of a weak pulse ($I=10^9$ W/cm$^2$) at the incident angle of 60$^\circ$,
(b) a strong pulse ($I=10^{13}$ W/cm$^2$) at the incident angle of 60$^\circ$,
(c) a weak pulse ($I=10^9$ W/cm$^2$) at the incident angle of 76$^\circ$,
and (d) a strong pulse ($I=10^{13}$ W/cm$^2$) at the incident angle of 76$^\circ$. 
We note that the incident angle of 76$^\circ$ corresponds to the Brewster's angle of Si with the index of refraction $n=4.0$ in the ALDA calculation.

In Figs.~\ref{fig:profile}(a) and (b), there appear both transmitted and reflected waves.
We find that the transmitted intensity is much lower in (b) than in (a).
Since we set the central frequency of the incident pulse at $\hbar \omega = 1.55$~eV, which is below the direct bandgap of Si, no absorption takes place at the weak intensity
shown in Fig.~\ref{fig:profile}(a).
At the higher intensity shown in Fig.~\ref{fig:profile}(b),
on the other hand, substantial absorption takes place at the film by multiphoton absorption processes, making the transmission weaker at the high intensity.

In Fig.~\ref{fig:profile}(c), the reflected wave cannot be seen and the transmitted wave has 
almost the same magnitude as the incident wave, as expected for the incidence
of the weak pulse without absorption at Brewster's angle.
Figure~\ref{fig:profile}(d) also shows for incidence at the Brewster's angle but with an increased pulse intensity, $I = 10^{13}$~W/cm$^2$.
Here, the magnitude of the transmitted wave is smaller than that of the incident pulse owing to the multiphoton absorption, as in the case shown in Fig.~\ref{fig:profile}(b).
Looking carefully at Fig.~\ref{fig:profile}(d), an interference pattern between the incident and reflected waves can be seen in the area surrounded by the dashed-line circle.
It indicates that that is slight reflection at Brewster's angle owing to nonlinear light-matter interaction for the strong incident pulse.

\begin{figure}
\centering
\includegraphics[width=0.6\textwidth]{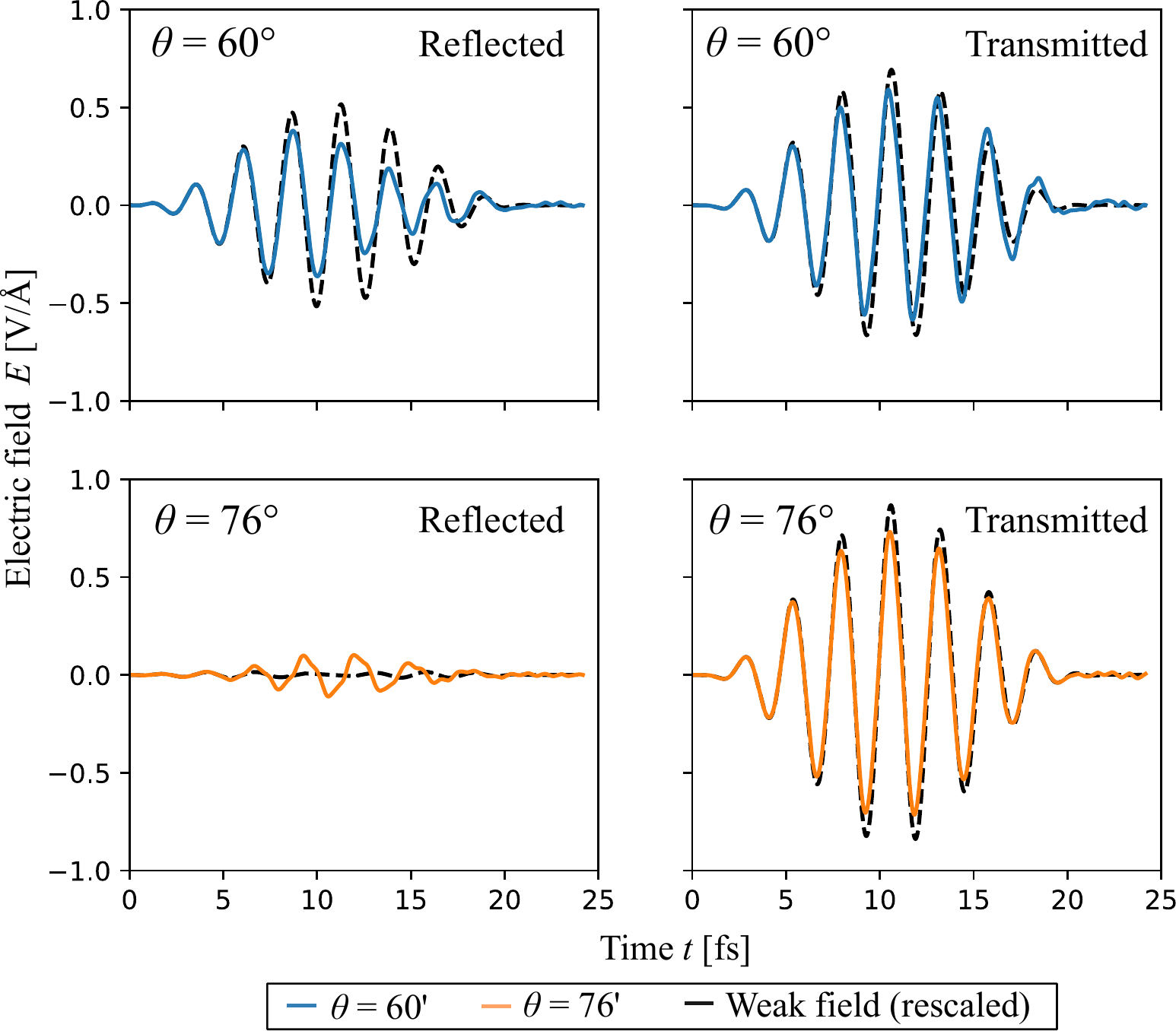}
\caption{ 
\label{fig:waveform2}
Reflected (left) and transmitted (right) pulse shapes in the multiscale Maxwell-TDDFT method.
Blue and orange curves correspond to the incident angles of $\theta=60^\circ$ and 76$^\circ$,
respectively, for a strong incident pulse of $I=10^{13}$ W/cm$^2$. Black-dashed lines are
for a weak pulse of $I=10^9$ W/cm$^2$ and are multiplied by a factor of 100.
}
\end{figure}

\begin{figure}[h!]
\centering
\includegraphics[width=0.6\textwidth]{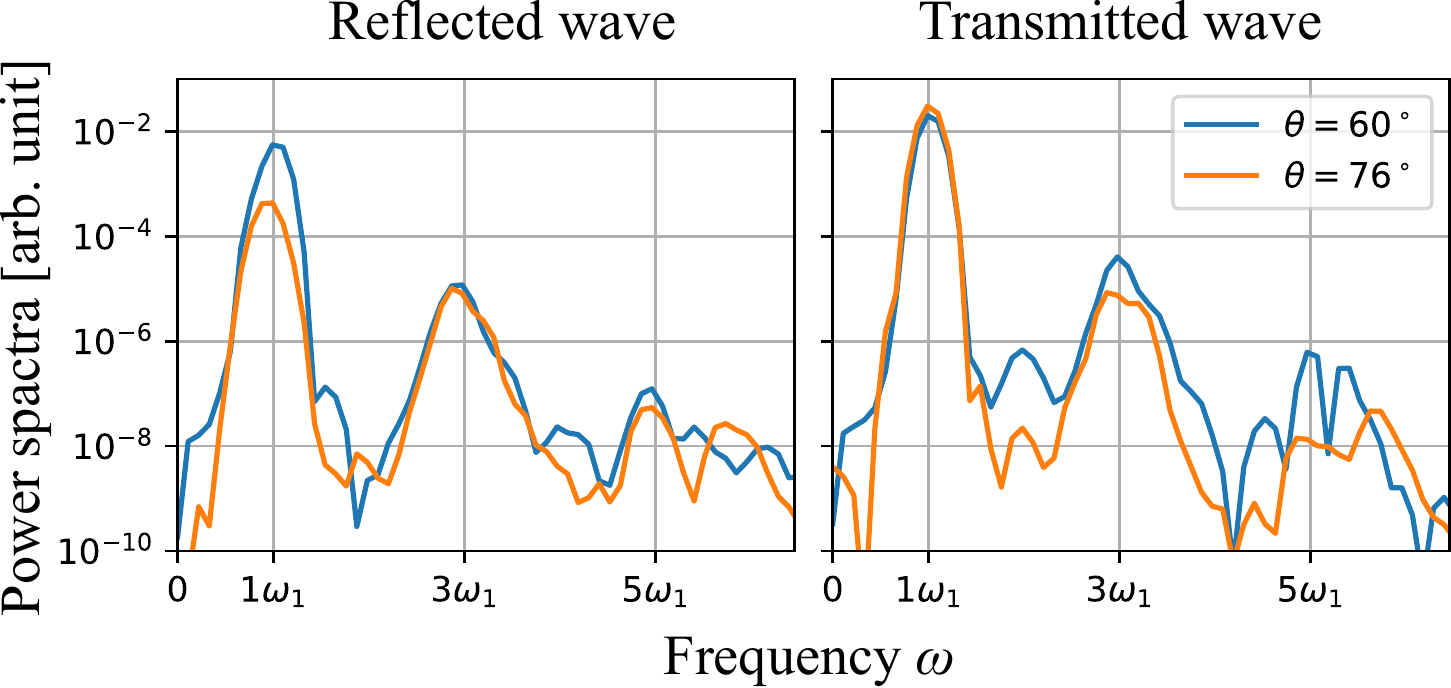}
\caption{
Spectral intensities of reflected (left) and transmitted (right) waves for incident angles of $\theta=60^\circ$ (blue) and 76$^\circ$ (orange).
\label{fig:power}
}
\end{figure}

To gain a deeper understanding of the propagation, waveforms of the reflected and transmitted pulses are shown in Fig. \ref{fig:waveform2} for the strong incident pulse of $I = 10^{13}$~W/cm$^2$ at the incident angles of 60$^\circ$ (blue) and 76$^\circ$ (orange), as well as waveforms for the weak pulse of $I=10^9$ W/cm$^2$ (black-dashed curve).
The latter is multiplied by a factor of 100 to show it in the same scale as the others.
At the angle of 60$^\circ$, the transmitted pulse looks similar between the strong and weak cases.
On the other hand, the reflected pulse in the strong case gradually becomes weaker with time.
This indicates that the optical property of the medium gradually changes with the continued irradiation of the strong pulse.
At Brewster's angle of 76$^\circ$, the reflected wave does not appear in the weak case, as expected.
However, a substantial reflection appears in the strong case. 
The waveform of the reflected wave also shows angled structures, indicating that it includes substantial nonlinear components.
For the transmitted wave, on the other hand, the amount of modulation looks rather small.
We thus find that nonlinear effects appear more clearly in the pulse shape of the reflected wave than in that of the  transmitted wave.

Figure~\ref{fig:power} shows power spectra of the reflected and transmitted waves. 
We find that the spectra include 3rd and 5th harmonic generations as well as the fundamental wave.
In the spectrum of the reflected wave, the fundamental wave component at Brewster's angle is one order of magnitude smaller than that at $60^\circ$. However, the magnitude of the nonlinear components is similar between $60^\circ$ and $76^\circ$. 
This finding explains why the reflected wave of the strong incident pulse at the Brewster's angle shows angled structures.
In the spectrum of the transmitted pulse, it is found that the nonlinear component at Brewster's angle is smaller than that at $60^\circ$.
This finding indicates that nonlinear effects show complex dependence on the incident angle, and the usefulness of the present approach when investigating them.

\begin{figure}[h!]
\centering
\includegraphics[width=0.65\textwidth]{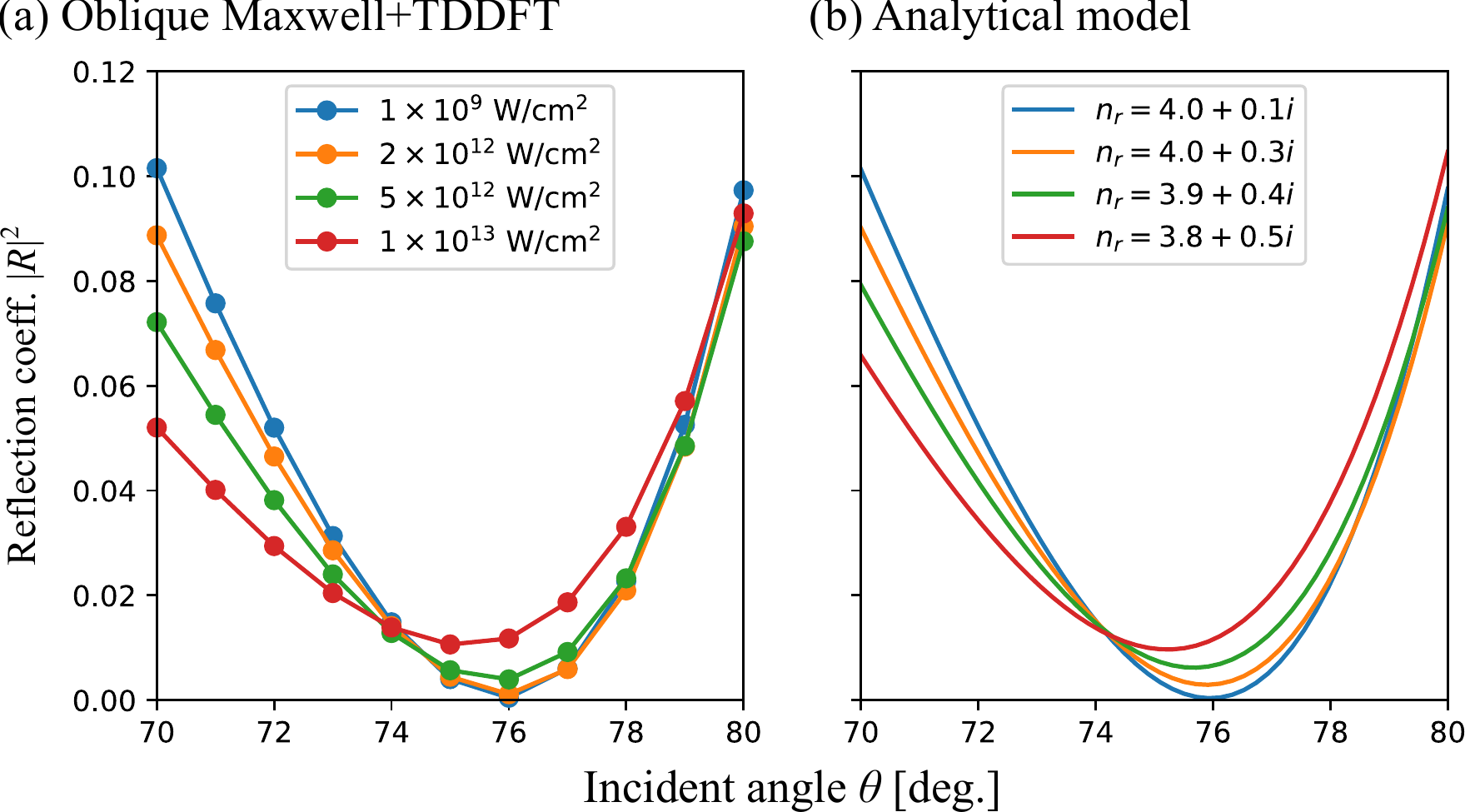}
\caption{ 
\label{fig:angle}
Angular dependence of the reflectance of a 50-nm-thick Si thin film.
(a)
Multiscale Maxwell+TDDFT calculation for four incident intensities from $10^9$ to $10^{13}$ W/cm$^2$.
(b)
Fresnel reflection using a complex dielectric constant determined to fit the reflectance of the
multiscale Maxwell-TDDFT calculation.}
\end{figure}

Finally, in Fig.~\ref{fig:angle}(a), we show a reflectance for incident pulses of various intensity and angles
in the multiscale Maxwell-TDDFT calculation.
For a weak pulse, the reflectance shows a clear minimum with zero reflectance at Brewster's angle.
As the intensities increases, the minimum value of the reflectance gradually increases.
The angle at the minimum reflectance also decreases.
In Fig.~\ref{fig:angle}(b), we show reflectance using the ordinary Fresnel formula with a complex
index of refraction. The value of the index of refraction at each intensity is determined by fitting the
result shown in Fig.~\ref{fig:angle}(a).
The fitted values of the index of refraction are indicated in the figure.
As seen from the figure, the behavior of the reflectance calculated by the multiscale Maxwell-TDDFT method
can be reasonably fitted by the complex index of refraction.
The imaginary part of the index of refraction determined by the fitting procedure gradually increases as the pulse intensity increases, while the real part decreases.

The change in the index of refraction may be understood in terms of the increase of the carrier density by the multiphoton excitation process. In the TDDFT calculation with a pump-probe set up, it has been shown that the excited carriers behave as free carriers \color{blue} and it adds the Drude-like term $-\sqrt{n_\mathrm{ex}}/(\omega^2+i\gamma\omega)$ to the dielectric function $\epsilon$ \cite{sato2014pumpprobe} where $\gamma$ is a decay constant. As the pulse intensity increases, the carrier density $n_\mathrm{ex}$ increases. Then the real part of the index of refraction $n_r$ ($=\sqrt{\epsilon}$) decreases, and the imaginary part increases. These behaviors of the reflectance are consistent with the results shown in Fig.~\ref{fig:angle}(a).
\color{black}

\section{\label{sec:summary} Summary}

We presented a theoretical formalism and numerical method to calculate nonlinear light propagation at oblique incidence based on time-dependent density functional theory (TDDFT), a first-principles computational method for electronic motion, coupled with the macroscopic Maxwell equations.
In our method, microscopic electronic motion is calculated by solving the time-dependent Kohn-Sham equation, the basic equation of TDDFT, at each grid point that is used to solve the macroscopic Maxwell equations.
Since the TDDFT calculation is computationally expensive, the number of grid points should be kept as small as possible. 
In our implementation, we express the Maxwell equations at oblique incidence as a one-dimensional wave equation for the vector potential and employ one-dimensional grid points to solve it.
In this way, the computational cost of the light propagation calculation at oblique incidence is similar to that of 
the light propagation at normal incidence. Since the normal component of the vector potential is discontinuous 
at the interface between the medium and the vacuum, we is effective proposed a smearing method to carry out stable
calculations and confirmed that it is effective.

As a demonstration of the method, we calculated the light propagation at oblique incidence on a silicon thin film.
It was shown that the multiscale Maxwell-TDDFT calculation adequately describes the linear optical response such as the absence of the reflected wave at Brewster's angle. 
For a strong incident pulse, the absence of reflectance at Brewster's angle is lifted by the nonlinear optical response.
It was shown that the angle dependence of the reflectance for strong incident pulses can be 
reasonably described by the nonlinear change in the index of refraction that is caused by multiphoton excitation.

\begin{backmatter}
\bmsection{Funding}
This work was financially supported by MEXT as a priority issue
theme 7 to be tackled by using Post-K Computer, and JST- CREST under Grant No. JP-MJCR16N5 and by JSPS KAKENHI under Grant Nos. 20K15194 and 20H02649.  
In addition, this work was also supported by Iketani Science and the Technology Foundation and  MEXT as "Program for Promoting Researches on the Supercomputer Fugaku" (Quantum-Theory-Based Multiscale Simulations toward the Development of Next-Generation Energy-Saving Semiconductor Devices, JPMXP1020200205).
Calculations were carried out at Oakforest-PACS at JCAHPC through the Multidisciplinary Cooperative Research Program in CCS, University of Tsukuba "First-principles electron-dynamics calculation for the interaction between the metasurface and extreme laser pulse", and through the HPCI System Research Project (Project No. hp190193).

\bmsection{Acknowledgments}
The authors would like thank to Dr. S. Yamada, Dr. A. Yamada and Dr. T. Takeuchi in Univ. of Tsukuba for the fruitful discussions. M. U. also thanks to Prof. T. Ono in Kobe Univ.

\bmsection{Disclosures}
The authors declare that there are no conflicts of interest related to this article

\bmsection{Data availability}
Data underlying the results presented in this paper are not publicly available at this time but may
be obtained from the authors upon reasonable request.

\medskip


\end{backmatter}

\bibliography{references}

\begin{thebibliography}{10}
\newcommand{\enquote}[1]{``#1''}

\bibitem{hecht2002optics}
E.~Hecht, \emph{Optics} (Pearson Education Limited, 2002).

\bibitem{gwon2010spectral}
H.~R. Gwon and S.~H. Lee, \enquote{Spectral and angular responses of surface
  plasmon resonance based on the kretschmann prism configuration,}
  {\protect\JournalTitle{Materials transactions}} \textbf{51}, 1150--1155
  (2010).

\bibitem{boyd2003nonlinear}
R.~W. Boyd, \emph{Nonlinear optics} (Elsevier, 2003).

\bibitem{schultze2013}
M.~Schultze, E.~M. Bothschafter, A.~Sommer, S.~Holzner, W.~Schweinberger,
  M.~Fiess, M.~Hofstetter, R.~Kienberger, V.~Apalkov, V.~S. Yakovlev, M.~I.
  Stockman, and F.~Krausz, \enquote{Controlling dielectrics with the electric
  field of light,} {\protect\JournalTitle{Nature}} \textbf{493}, 75--78 (2013).

\bibitem{lucchini2016attosecond}
M.~Lucchini, S.~A. Sato, A.~Ludwig, J.~Herrmann, M.~Volkov, L.~Kasmi,
  Y.~Shinohara, K.~Yabana, L.~Gallmann, and U.~Keller, \enquote{Attosecond
  dynamical franz-keldysh effect in polycrystalline diamond,}
  {\protect\JournalTitle{Science}} \textbf{353}, 916--919 (2016).

\bibitem{sommer2016attosecond}
A.~Sommer, E.~Bothschafter, S.~Sato, C.~Jakubeit, T.~Latka, O.~Razskazovskaya,
  H.~Fattahi, M.~Jobst, W.~Schweinberger, V.~Shirvanyan \emph{et~al.},
  \enquote{Attosecond nonlinear polarization and light--matter energy transfer
  in solids,} {\protect\JournalTitle{Nature}} \textbf{534}, 86--90 (2016).

\bibitem{ciappina2017}
M.~F. Ciappina, J.~A. Perez-Hernandez, A.~S. Landsman, W.~A. Okell,
  S.~Zherebtsov, B.~Foerg, J.~Schoetz, L.~Seiffert, T.~Fennel, T.~Shaaran,
  T.~Zimmermann, A.~Chacon, R.~Guichard, A.~Zaier, J.~W.~G. Tisch, J.~P.
  Marangos, T.~Witting, A.~Braun, S.~A. Maier, L.~Roso, M.~Krueger,
  P.~Hommelhoff, M.~F. Kling, F.~Krausz, and M.~Lewenstein,
  \enquote{{Attosecond physics at the nanoscale},}
  {\protect\JournalTitle{{REPORTS ON PROGRESS IN PHYSICS}}} \textbf{{80}}
  ({2017}).

\bibitem{vampa2018hhgbackward}
G.~Vampa, Y.~S. You, H.~Liu, S.~Ghimire, and D.~A. Reis, \enquote{Observation
  of backward high-harmonic emission from solids,}
  {\protect\JournalTitle{OPTICS EXPRESS}} \textbf{26}, 12210--12218 (2018).

\bibitem{veysoglu1993}
M.~Veysoglu, R.~Shin, and J.~Kong, \enquote{A finite-difference time-domain
  analysis of wave scattering from periodic surfaces: Oblique incidence case,}
  {\protect\JournalTitle{Journal of Electromagnetic Waves and Applications}}
  \textbf{7}, 1595--1607 (1993).

\bibitem{roden1998}
J.~Roden, S.~Gedney, M.~Kesler, J.~Maloney, and P.~Harms, \enquote{Time-domain
  analysis of periodic structures at oblique incidence: orthogonal and
  nonorthogonal fdtd implementations,} {\protect\JournalTitle{IEEE Transactions
  on Microwave Theory and Techniques}} \textbf{46}, 420--427 (1998).

\bibitem{harms1994implementation}
P.~Harms, R.~Mittra, and W.~Ko, \enquote{Implementation of the periodic
  boundary condition in the finite-difference time-domain algorithm for fss
  structures,} {\protect\JournalTitle{IEEE Transactions on Antennas and
  Propagation}} \textbf{42}, 1317--1324 (1994).

\bibitem{baida2009split}
F.~Baida and A.~Belkhir, \enquote{Split-field fdtd method for oblique incidence
  study of periodic dispersive metallic structures,}
  {\protect\JournalTitle{Optics letters}} \textbf{34}, 2453--2455 (2009).

\bibitem{valuev2008iterative}
I.~Valuev, A.~Deinega, and S.~Belousov, \enquote{Iterative technique for
  analysis of periodic structures at oblique incidence in the finite-difference
  time-domain method,} {\protect\JournalTitle{Optics letters}} \textbf{33},
  1491--1493 (2008).

\bibitem{varin2018}
C.~Varin, R.~Emms, G.~Bart, T.~Fennel, and T.~Brabec, \enquote{Explicit
  formulation of second and third order optical nonlinearity in the fdtd
  framework,} {\protect\JournalTitle{Computer Physics Communications}}
  \textbf{222}, 70--83 (2018).

\bibitem{kira2011sbe}
M.~Kira and S.~W. Koch, \emph{Semiconductor Quantum Optics} (Cambridge
  University Press, 2011).

\bibitem{yabana2012time}
K.~Yabana, T.~Sugiyama, Y.~Shinohara, T.~Otobe, and G.~Bertsch,
  \enquote{Time-dependent density functional theory for strong electromagnetic
  fields in crystalline solids,} {\protect\JournalTitle{Physical Review B}}
  \textbf{85}, 045134 (2012).

\bibitem{floss2018hhg}
I.~Floss, C.~Lemell, G.~Wachter, V.~Smejkal, S.~A. Sato, X.-M. Tong, K.~Yabana,
  and J.~Burgd\"orfer, \enquote{Ab initio multiscale simulation of high-order
  harmonic generation in solids,} {\protect\JournalTitle{Phys. Rev. A}}
  \textbf{97}, 011401 (2018).

\bibitem{sato2015propagation}
S.~A. Sato, K.~Yabana, Y.~Shinohara, T.~Otobe, K.-M. Lee, and G.~F. Bertsch,
  \enquote{Time-dependent density functional theory of high-intensity
  short-pulse laser irradiation on insulators,} {\protect\JournalTitle{Phys.
  Rev. B}} \textbf{92}, 205413 (2015).

\bibitem{zhang2010}
L.~Zhang and T.~Seideman, \enquote{Rigorous formulation of oblique incidence
  scattering from dispersive media,} {\protect\JournalTitle{Phys. Rev. B}}
  \textbf{82}, 155117 (2010).

\bibitem{liu2018plasma}
J.-X. Liu, Z.-K. Yang, L.~Ju, L.-Q. Pan, Z.-G. Xu, and H.-W. Yang,
  \enquote{Boltzmann finite-difference time-domain method research
  electromagnetic wave oblique incidence into plasma,}
  {\protect\JournalTitle{Plasmonics}} \textbf{13}, 1699--1704 (2018).

\bibitem{runge1984density}
E.~Runge and E.~K. Gross, \enquote{Density-functional theory for time-dependent
  systems,} {\protect\JournalTitle{Physical Review Letters}} \textbf{52}, 997
  (1984).

\bibitem{yabana1996time}
K.~Yabana and G.~Bertsch, \enquote{Time-dependent local-density approximation
  in real time,} {\protect\JournalTitle{Physical Review B}} \textbf{54}, 4484
  (1996).

\bibitem{bertsch2000}
G.~F. Bertsch, J.-I. Iwata, A.~Rubio, and K.~Yabana, \enquote{Real-space,
  real-time method for the dielectric function,} {\protect\JournalTitle{Phys.
  Rev. B}} \textbf{62}, 7998--8002 (2000).

\bibitem{uemoto2019nonlinear}
M.~Uemoto, Y.~Kuwabara, S.~A. Sato, and K.~Yabana, \enquote{Nonlinear
  polarization evolution using time-dependent density functional theory,}
  {\protect\JournalTitle{The Journal of chemical physics}} \textbf{150}, 094101
  (2019).

\bibitem{otobe2012hhg}
T.~Otobe, \enquote{First-principle description for the high-harmonic generation
  in a diamond by intense short laser pulse,} {\protect\JournalTitle{Journal of
  Applied Physics}} \textbf{111}, 093112 (2012).

\bibitem{yamada2019energy}
A.~Yamada and K.~Yabana, \enquote{Energy transfer from intense laser pulse to
  dielectrics in time-dependent density functional theory,}
  {\protect\JournalTitle{The European Physical Journal D}} \textbf{73}, 87
  (2019).

\bibitem{uemoto2021graphite}
M.~Uemoto, S.~Kurata, N.~Kawaguchi, and K.~Yabana, \enquote{First-principles
  study of ultrafast and nonlinear optical properties of graphite thin films,}
  {\protect\JournalTitle{Phys. Rev. B}} \textbf{103}, 085433 (2021).

\bibitem{yamada2019cp}
A.~Yamada and K.~Yabana, \enquote{Multiscale time-dependent density functional
  theory for a unified description of ultrafast dynamics: Pulsed light,
  electron, and lattice motions in crystalline solids,}
  {\protect\JournalTitle{Phys. Rev. B}} \textbf{99}, 245103 (2019).

\bibitem{ghosh1998handbook}
G.~Ghosh, \emph{Handbook of optical constants of solids: Handbook of
  thermo-optic coefficients of optical materials with applications} (Academic
  Press, 1998).

\bibitem{noda2019salmon}
M.~Noda, S.~A. Sato, Y.~Hirokawa, M.~Uemoto, T.~Takeuchi, S.~Yamada, A.~Yamada,
  Y.~Shinohara, M.~Yamaguchi, K.~Iida \emph{et~al.}, \enquote{Salmon: Scalable
  ab-initio light--matter simulator for optics and nanoscience,}
  {\protect\JournalTitle{Computer Physics Communications}} \textbf{235},
  356--365 (2019).

\bibitem{troullier1993}
N.~Troullier and J.~L. Martins, \enquote{Efficient pseudopotentials for
  plane-wave calculations,} {\protect\JournalTitle{Phys. Rev. B}} \textbf{43},
  1993--2006 (1991).

\bibitem{kleinman1982}
L.~Kleinman and D.~M. Bylander, \enquote{Efficacious form for model
  pseudopotentials,} {\protect\JournalTitle{Phys. Rev. Lett.}} \textbf{48},
  1425--1428 (1982).

\bibitem{perdew1981}
J.~P. Perdew and A.~Zunger, \enquote{Self-interaction correction to
  density-functional approximations for many-electron systems,}
  {\protect\JournalTitle{Phys. Rev. B}} \textbf{23}, 5048--5079 (1981).

\bibitem{sato2014pumpprobe}
S.~A. Sato, K.~Yabana, Y.~Shinohara, T.~Otobe, and G.~F. Bertsch,
  \enquote{Numerical pump-probe experiments of laser-excited silicon in
  nonequilibrium phase,} {\protect\JournalTitle{Phys. Rev. B}} \textbf{89},
  064304 (2014).

\end{thebibliography}

\end{document}